\documentclass[a4paper,11pt]{article}
\pdfoutput=1 

\usepackage{jinstpub} 
\usepackage{palatino}
\usepackage{enumitem}
\newcommand{\boro}{B$_4$C\:}
\newcommand{\litio}{$^7$Li}
\newcommand{\alfa}{$\alpha$}
\usepackage{lineno}

\title{\boldmath{u-RANIA: a neutron detector based on \textmu -RWELL technology}}

\author[a,b,1]{I.~Balossino,\note{Corresponding author.}}
\author[c]{G.~Bencivenni,}
\author[i]{P.~Bielowka,}
\author[a]{G.~Cibinetto,}
\author[a]{R.~Farinelli,}
\author[c]{G.~Felici,}
\author[a]{I.~Garzia,}
\author[c]{M.~Gatta,}
\author[e]{P.~Giacomelli,}
\author[c]{M.~Giovannetti,}
\author[g]{R.~Hall Wilton,}
\author[g]{C.-C.~Lai,}
\author[d]{L.~Lavezzi,}
\author[f]{F.~Messi,}
\author[a,b]{G.~Mezzadri}
\author[c]{G.~Morello,}
\author[h]{M.~Pinamonti,}
\author[c]{M.~Poli Lener,}
\author[g]{L.~Robinson,}
\author[a]{M.~Scodeggio,}
\author[g]{P.-O.~Svensson}

\affiliation[a]{INFN-Ferrara,\\via Saragat 1, 44122 Ferrara, Italy} 
\affiliation[b]{Institute of High Energy Physics, Chinese Academy of Sciences,\\ 19B Yuquan Road, Shijingshan District, 100049 Beijing, China}
\affiliation[c]{INFN-National Laboratory of Frascati,\\ via Enrico Fermi 54, 00044 Frascati (Roma) Italy}
\affiliation[d]{INFN-Torino,\\via P.Giuria 1, 10125 Torino, Italy}
\affiliation[e]{INFN-Bologna,\\viale Berti Pichat 6/2, 40127 Bologna, Italy}
\affiliation[f]{Lund University,\\Box 117, 221 00 Lund, Sweden}
\affiliation[g]{Detector Group, European Spallation Source ERIC (ESS)\\Box 176, SE-221 00 Lund, Sweden}
\affiliation[h]{ELTOS SPA\\Strada E 44, San Zeno, 52100 Arezzo, Italy}
\affiliation[i]{TECHTRA S.p.z.o.o.\\ul. Dunska 13 54-427 Wroclaw, Polska}

\emailAdd{balossino@fe.infn.it}

\abstract{In the framework of the ATTRACT-uRANIA project, funded by the European Community, we are developing an innovative neutron imaging detector based on micro-Resistive WELL (\textmu-RWELL) technology. 
The \textmu-RWELL, based on the resistive detector concept, ensuring an efficient spark quenching mechanism, is a highly reliable device. It is composed by two main elements: a readout-PCB and a cathode. The amplification stage for this device is embedded in the readout board through a resistive layer realized by means of an industrial process with DLC (Diamond-Like Carbon). A thin layer of \boro on the copper surface of the catode allows the thermal neutrons detection through the release of \litio\: and \alfa\: particles in the active volume. This technology has been developed to be an efficient and convenient alternative to the $^3$He shortage. The goal of the project is to prove the feasibility of such a novel neutron detector by developing and testing small planar prototypes with readout boards suitably segmented with strip or pad readout, equipped with existing electronics or readout in current mode.
Preliminary results from the test with different prototypes, showing a good agreement with the simulation, will be presented together with construction details of the prototypes and the future steps of the project.

}


\proceeding{International Conference Instrumentation for Colliding Beam Physics\\
  24-28 February 2020\\
  Budker Institute of Nuclear Physics, Novosibirsk, Russia}

\begin{document}
\maketitle

\section{The Project}\label{sec:intro}

The working framework is the ATTRACT project funded by the European Community. In particular, the uRANIA - micro Resistive Advanced Neutron Imaging Apparatus - proposal foresees the development of an innovative detector based on micro-resistive well technology~\cite{a} to perform neutron imaging. Such technology, based on resistive detector concept, is highly reliable and ensures an efficient spark quenching mechanism. The goal of this project is to develop, build and test planar prototypes and, therefore, prove its feasibility. 

\begin{figure}[t!]
\centering 
\includegraphics[width=0.6\textwidth,clip]{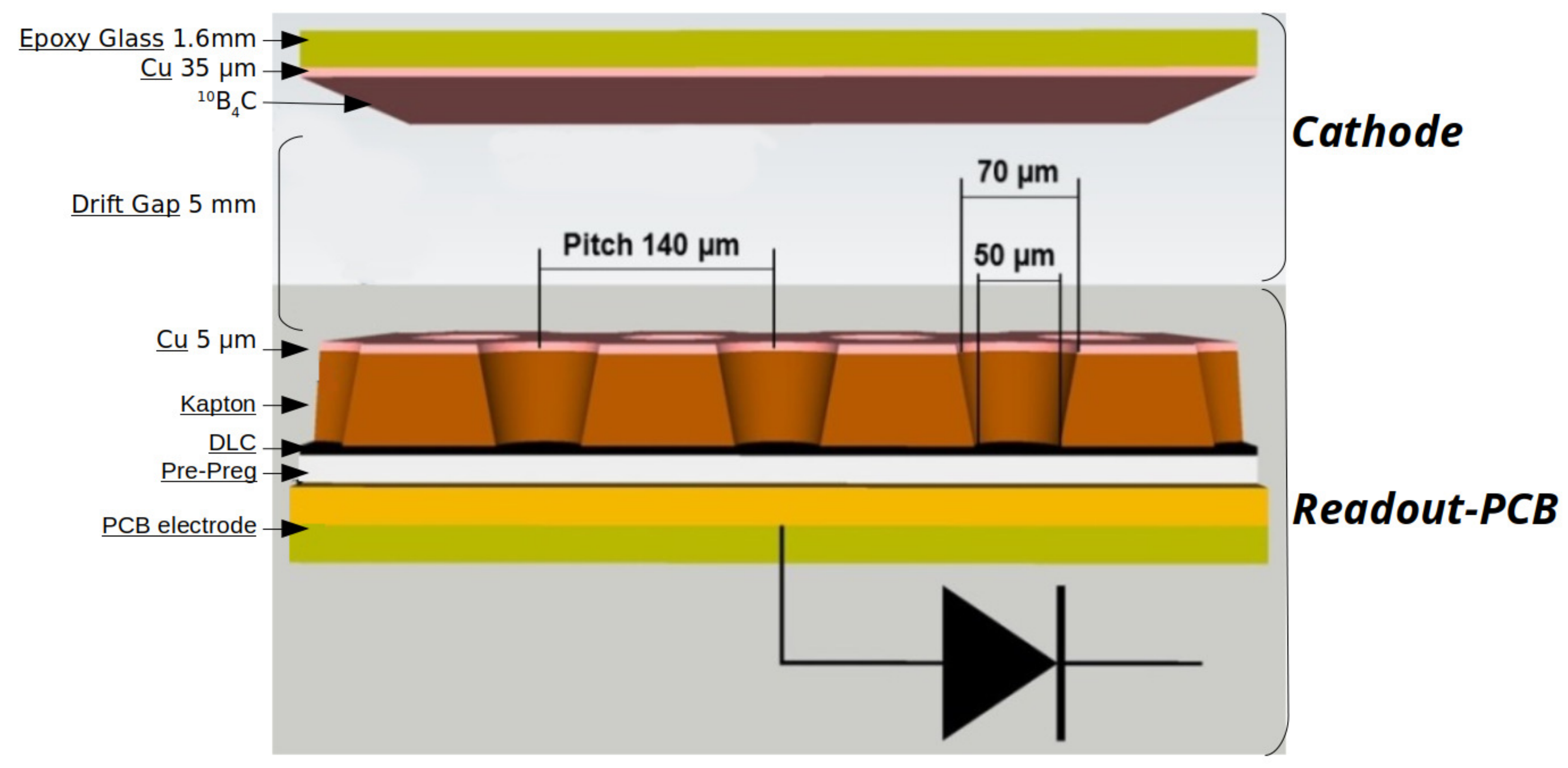}
\caption{\label{fig:1} \textmu-RWELL design}
\end{figure}

The device is composed by two main elements as shown in figure~\ref{fig:1}:\vspace{0.01pt}
\begin{description}[style=unboxed,leftmargin=0cm,itemsep=0.1pt,parsep=0.1pt]
\item [Cathode] A standard \textmu-RWELL cathode PCB is composed by a Cu layer on a glass epoxy plate. In this case, it is coated with few \textmu m of \boro deposition. Such deposition is performed by the ESS Coating Workshop in Link\"{o}ping, Sweden~\cite{b}. This \boro layer is the crucial part for the conversion of the thermal neutrons in detectable particles. Such kind of technology is being developed to be as an efficient and convenient alternative to the $^{3}$He shortage.  Four prototypes have been produced with different thicknesses of \boro to validate the simulations and to study the optimal value to maximize the performances. The samples thicknesses are 1.5, 2.5, 3.5 and 4.5 \textmu m.
\item [Readout-PCB] This part includes the anode together with the amplification stage: a WELL patterned Apical\textregistered \: foil. This is realized by photolithography as a matrix of wells (diameter of 60-70 \textmu m; pitch of 140 \textmu m) on a 50 \textmu m thick polymide substrate by the TECHTRA agency, Poland~\cite{c}. It is then embedded in the readout board - prepared by the ELTOS S.p.A., Italy~\cite{d} - through a resistive layer realized by means of an industrial process with Diamond-Like Carbon (DLC)~\cite{e}. The resistivity of this layer, whose typical range is from few tens to hundreds of MOhm/$\square$, must be optimized to maximize the detector performances in terms of rate capability, spark amplitude quenching and maximum available gain. 
The readout boards will be suitably segmented with strip or pad readout, optimized to be equipped with already existing MPGD electronics. 
\end{description}

The detection of thermal neutrons (kinetic energy of about 0.025 eV) will be performed thanks to the release of \litio\: and \alfa\: particles inside the active volume. The \boro on the copper surface of the cathode is the key of this process. The reactions that take place are:
\begin{subequations}\label{eq:y}
\begin{align}
\label{eq:1}
n + ^{10}_{5}\mathrm{B} \rightarrow & \: ^{7}_{3}\mathrm{Li}~(0.84~\mathrm{MeV}) + \alpha~(1.47~\mathrm{MeV}) + \gamma~(0.48~\mathrm{MeV})
\\
\label{eq:2}
n + ^{10}_{5}\mathrm{B} \rightarrow & \: ^{7}_{3}\mathrm{Li}~(1.02~\mathrm{MeV}) + \alpha~(1.78~\mathrm{MeV})
\end{align}
\end{subequations}
where cases presented in eq.~\ref{eq:1} takes places the $94\%$ of the times and the remaining $6\%$ involves eq.~\ref{eq:2}. In both equations, the particle energies at emission in boron rest frame are reported. The \litio/\alfa\: production happens back to back and this allows to have mutually exclusive events. The conversion results have non-negligible cross section with the \boro-coated cathode.  The four prototypes with different thicknesses of the \boro layer are considered to investigate this aspect and to understand which is the best thickness to be sure that the conversion results reach the gas gap.

\section{The Beam Test}

\begin{figure}[b!]
\centering 
\includegraphics[width=0.4\textwidth,clip]{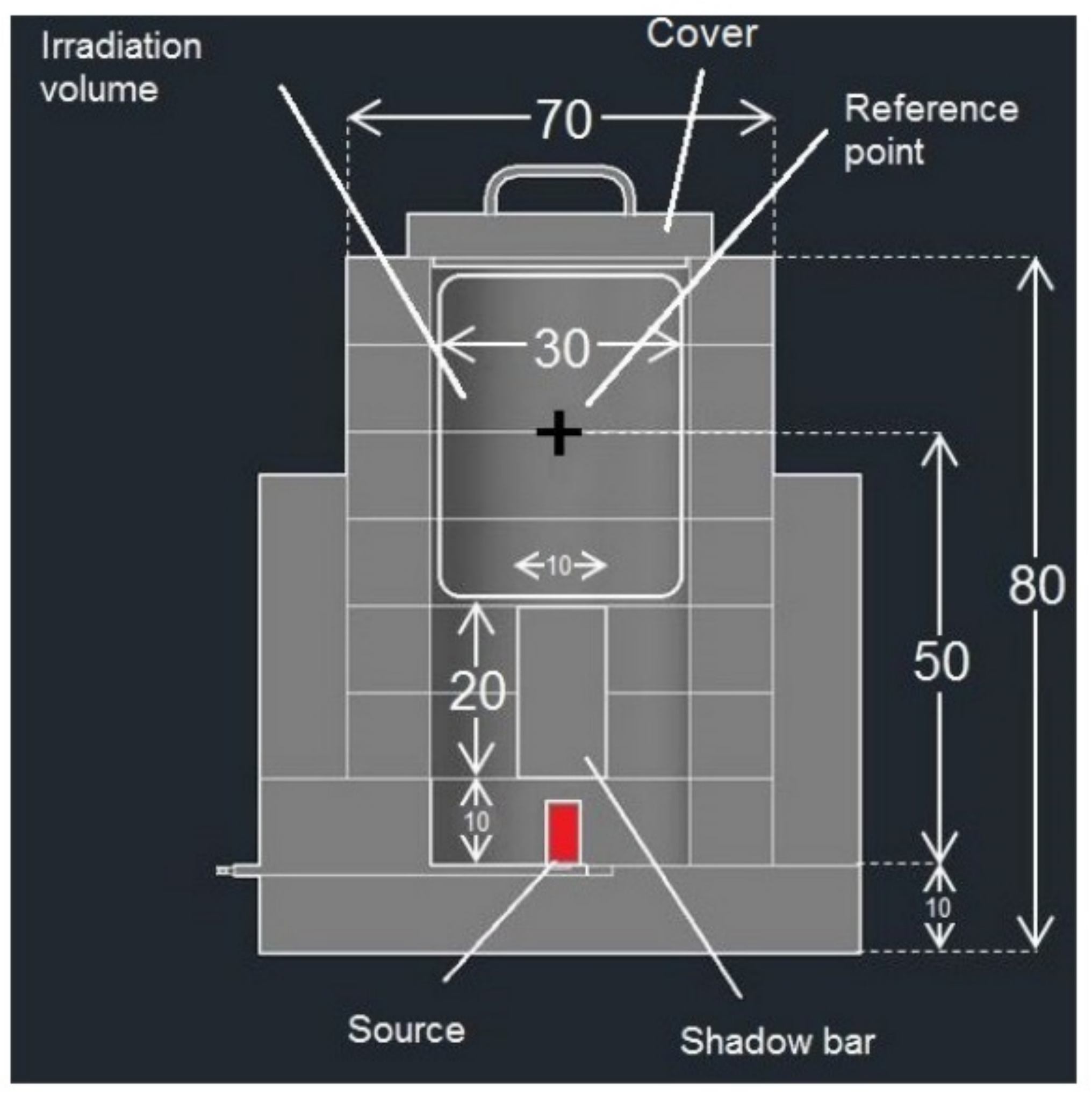}
\caption{\label{fig:2} HOTNESS design; the measurements are in cm}
\end{figure}

A preliminary characterization of the prototypes has been performed at the ENEA-HOTNES in Frascati~\cite{f}. The detectors have been exposed to a thermal neutron source, operating with the gas mixture Ar/CO$_2$/CF$_4$ (45/15/40) and read out in current mode to measure the neutron detection efficiency.
\paragraph{Facility}
The Frascati facility offers an HOmogeneus Thermal Neutron Source (HOTNES) shown in figure~\ref{fig:2}.
The source is $^{241}$Am-B and it is placed at the bottom of the cylindrical cavity delimited by polyethylene walls. On top of it, a shadow bar prevents fast neutrons to directly reach the samples under test. Inside the irradiation volume, when the cover is in place, the fluence is almost uniform and it is fully characterized for thermal neutrons. At the reference point, in the middle of the volume, the thermal neutrons fluence is expected to be $\mathrm{758 \pm 16}$ Hz/cm$^2$ with very low gamma background.

\paragraph{Calibration}
To extract from the current measurement the neutron detection efficiency, a detailed simulation of the thermal neutrons detection process as well as a fine calibration of each detector has been performed.

As preparation for the test, the devices have been calibrated in gain as a function of the applied high voltage, reproducing the beam test conditions by means of an X-ray gun. Testing at high voltage range between 300 and 700 V, the detector gain increases exponentially from 10$^2$ to 10$^4$.

At the facility, a first reference test has been performed with a cathode without the boron coating to check the background inside HOTNESS. An increase of the current drained by the resistive layer has been reported, but it is negligible with respect to the operating point. 

The current drain is also investigated for all the devices as a function of the gain when large ionization occurs (<N$^\alpha_{primary}$> = 2.2 $\cdot$ 10$^{4}$; <N$^{\mathrm{Li}}_{primary}$> = 1.1 $\cdot$ 10$^{4}$). At lower gains, the current absorption linearly increases, while at higher gains some rate effect occurs. It has been decided to proceed for the test in the gain range between 200 and 700. 

Another test performed before reaching for the final results, involves the attenuation of the current drain due to the FR-4 epoxy glass of the support layer used for the cathode. The current drain has been investigated by stacking an increasing number of cathodes together, all without \boro coating. The results confirm the simulations performed with PCB composed by boron that absorbes the 10$\%$ of the neutrons and hydrogen that scatteres the 9$\%$. Every cathode decreases the current drain by (17.7 $\pm$ 0.7)$\%$. If the expected nominal flux is 758 $\pm$ 16 Hz/cm$^2$ such cathode effect brings the effective flux to 624 $\pm$ 28 Hz/cm$^2$.

\paragraph{Results}
The detection efficiency of such prototypes has been simulated to be compared with the measured results. 
The simulation has been performed calculating the sum of \litio\: and \alfa\: particles that reach the gas over the total number of neutrons produced. 
The measured efficiency is calculated as follows:

\begin{equation}\label{eq:3}
\epsilon = \frac{I}{e \cdot R \cdot G \cdot <N>}
\end{equation}

where:
\begin{itemize}\itemsep0.01pt
\item \textit{I} is the current measured from the resistive;
\item \textit{e} is the electron charge;
\item \textit{R} is the thermal neutrons fluence irradiating the detector;
\item \textit{G} is the gain of the detector;
\item \textit{<N>} is the average ionization generated in the gas gap by the \alfa\: or \litio\: ions coming from the interaction of the thermal neutrons with boron atoms, estimated by the simulations.
\end{itemize}

\begin{figure}[htbp]
\centering 
\includegraphics[width=0.6\textwidth,clip]{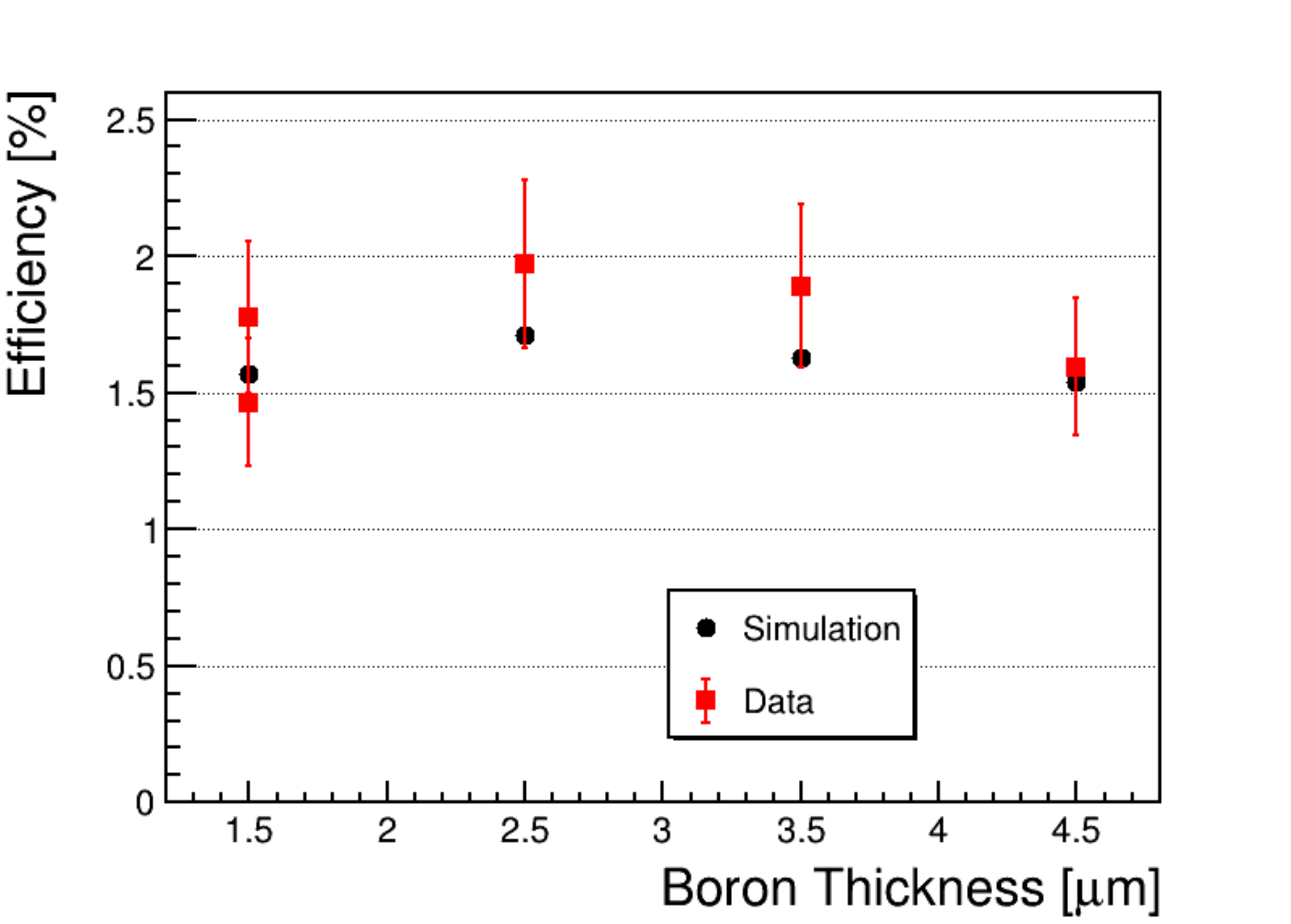}
\caption{\label{fig:3} Efficiency wrt \boro coating thickness}
\end{figure}

The results of the efficiency are reported in figure~\ref{fig:3} with respect to the thickness of the \boro coating on the cathode ranging from $1.5 \div 4.5$ \textmu m. The results reported are the measured data with the nominal flux (red square) and the simulations (black dot). It is possible to see that there is correlation between all the results and an overall neutron detection efficiency ranging between $1.5 \div 2.0\,\%$ for the nominal flux is found.

\section{The Future}

Some preliminary tests have been perfomed on four prototypes of an innovative detector build with micro-resistive well technology to perform neutron imaging. Such devices foresee the detection of thermal neutrons through their reaction with a thin layer of \boro on the cathode that results in the production of \litio\: and \alfa\: particles. Each prototype has been built with different thicknesses of the \boro layer:  1.5, 2.5, 3.5 and 4.5 \textmu m. 

In this preliminary beam test, the measurements performed demonstrate the possibility to reach a neutron detection efficiency ranging between $1.5 \div 2\,\%$ within all the four detectors.

Several systematic effects due to the presence of FR-4 glass epoxy structure on the cathode PCB such as the absorption or back scattering of the thermal neutrons have been taken into account both in the measurements and in the simulation.

In order to improve the detection efficiency, it is planned to upgrade the detector with a boron-coated aluminum or copper mesh that will be placed between the cathode and the readout PCB. With this new layout of the detector a second beam test will be perfomed. 

The spatial resolution with such a detector is expected to be of the order of 100 microns. Such performance will be possible with a charge and time readout electronics inherited from the BESIII Cylindrical GEM detector~\cite{g}. Such readout electronics is able to process charge and time information and, together with the already developed charge centroid and \textmu TPC clasterization algorithms~\cite{h}, will be adapted to this new detector to allow the recontruction of the position of the neutron interaction. 

\acknowledgments

This work is supported by the Italian institute of nuclear physics (INFN).
This project has received funding from the ATTRACT project funded by the EC under Grant Agreement 777222.


\end{document}